\documentclass[%
 reprint,
superscriptaddress,
 amsmath,amssymb,longbibliography,
 aps,
pra,
]{revtex4-1}

\usepackage{setspace}
\usepackage{graphicx}
\usepackage{bbold}
\usepackage{amsmath}
\usepackage{amsfonts}
\usepackage{amsthm}
\usepackage{float}
\usepackage[margin=1in]{geometry}
\usepackage{enumerate}
\usepackage{etoolbox}
\usepackage{multirow}
\usepackage{amssymb}
\usepackage{comment}
\usepackage{mathrsfs}
\usepackage{ gensymb }
\usepackage{braket}
\usepackage{siunitx}
\usepackage{dcolumn}
\usepackage{subcaption}
\usepackage{xcolor}
\definecolor{dred}{rgb}{.8,0.2,.2}

\begin{document}

\newcommand{\moire}[0]{moir\'e\ }

\title{Optimization of neutron diffraction from phase-gratings}

\author{B. Heacock}
\email{bjheacoc@ncsu.edu}
\affiliation{Department of Physics, North Carolina State University, Raleigh, NC 27695, USA}
\affiliation{Triangle Universities Nuclear Laboratory, Durham, North Carolina 27708, USA}
\author{D. Sarenac}
\affiliation{Institute for Quantum Computing, University of Waterloo,  Waterloo, ON, Canada, N2L3G1}
\affiliation{Department of Physics, University of Waterloo, Waterloo, ON, Canada, N2L3G1}
\author{D. G. Cory}
\affiliation{Institute for Quantum Computing, University of Waterloo,  Waterloo, ON, Canada, N2L3G1} 
\affiliation{Department of Chemistry, University of Waterloo, Waterloo, ON, Canada, N2L3G1}
\affiliation{Perimeter Institute for Theoretical Physics, Waterloo, ON, Canada, N2L2Y5}
\affiliation{Canadian Institute for Advanced Research, Toronto, Ontario, Canada, M5G 1Z8}
\author{M. G. Huber}
\affiliation{National Institute of Standards and Technology, Gaithersburg, Maryland 20899, USA}
\author{D. S. Hussey}
\affiliation{National Institute of Standards and Technology, Gaithersburg, Maryland 20899, USA}
\author{C. Kapahi}
\affiliation{Institute for Quantum Computing, University of Waterloo,  Waterloo, ON, Canada, N2L3G1}
\affiliation{Department of Physics, University of Waterloo, Waterloo, ON, Canada, N2L3G1}
\author{H. Miao}
\affiliation{Biophysics and Biochemistry Center, National Heart, Lung and Blood Institute, National Institutes of Health, Bethesda, Maryland USA}
\author{H. Wen}
\affiliation{Biophysics and Biochemistry Center, National Heart, Lung and Blood Institute, National Institutes of Health, Bethesda, Maryland USA}
\author{D. A. Pushin}
\email{dmitry.pushin@uwaterloo.ca}
\affiliation{Institute for Quantum Computing, University of Waterloo,  Waterloo, ON, Canada, N2L3G1}
\affiliation{Department of Physics, University of Waterloo, Waterloo, ON, Canada, N2L3G1}

\begin{abstract}
The recent development of phase-grating moir\'e\ neutron interferometry promises a wide range of impactful experiments from dark-field imaging of material microstructure to precise measurements of fundamental constants. However, the contrast of 3 \% obtained using this moir\'e\ interferometer was well below the theoretical prediction of 30 \% using ideal gratings.  It is suspected that non-ideal aspects of the phase-gratings  was a leading contributor to this deficiency and that phase-gratings needed to be quantitatively assessed and optimized. 
Here we characterize neutron diffraction from phase-gratings using Bragg diffraction crystals to determine the optimal phase-grating orientations. We show well-defined diffraction peaks and explore perturbations to the diffraction peaks and the effects on interferometer contrast as a function of grating alignment.  This technique promises to improve the contrast of the grating interferometers by providing in-situ aides to grating alignment.
\end{abstract}

\maketitle

\section{Introduction}

Neutrons have played an important role in contemporary physics due to their relatively large mass, electric neutrality, and sub-nanometer-sized wavelengths. Numerous neutron techniques have been developed for the precise measurements of material properties and fundamental constants \cite{sears1989neutron}. In particular, the perfect-crystal neutron interferometer (NI) \cite{ni_book2ed} has made numerous impacts in exploring the properties of the neutron and its interactions \cite{oam,chameleon,decoherence,holography,Klepp_2014,Denkmayr2014}. Nonetheless, the wide spread adoption of this technique was hindered by its narrow wavelength acceptance and the stringent requirements for environmental isolation, beam collimation, and setup alignment \cite{bauspiess1978prototype,arif1993facilities,zawisky2010large,saggu2016decoupling}.


A less demanding setup is a grating-based, Mach-Zehnder interferometer which relies on the coherent control and mixing of the various grating diffraction orders \cite{Ioffe1985b,GRUBER1989363,VanDerZouw2000,Schellhorn1997,Klepp2011}. However, these interferometers require  using cold ($\lambda \sim 5 \; \mathrm{\AA}$) or very cold neutrons ($\lambda \sim 100 \; \mathrm{\AA}$) with a high degree of collimation; greatly limiting their wide spread use. Some of these limitations are overcome by Talbot-Lau grating interferometers which work in the full-field of a cold neutron beam, with relaxed requirements for wavelength acceptance \cite{Clauser1994,Pfeiffer2006,lee2009development}. Talbot-Lau grating interferometers have a peak visibility for the design wavelength and have the negative aspect of needing high aspect ratio neutron transmission gratings, which are a fabrication challenge.

The recent advances in far-field phase-grating \moire interferometry extends phase-grating neutron interferometry to  high intensity sources of thermal neutrons, showing promise for impactful long-path-length experiments and providing new methods of characterizing materials by scanning the autocorrelation length over many orders of magnitude \cite{twogratings,hussey2016demonstration,sarenac2018three,brooks2017neutron,brooks2018neutron}. This novel technique has proven to be advantageous over other neutron interferometry setups in terms of broader wavelength acceptance and less stringent alignment requirements. However, the observed contrast of 3 \% \cite{sarenac2018three} needs to be substantially improved.

A major contributor to the observed poor contrast is suspected to be the sub-optimal quality of the high aspect ratio phase-gratings, and a follow up tomography study confirmed that the shape of the phase-gratings that were used was not ideal \cite{heacock2018sub}. Here we explore direct methods of optimizing the performance of the individual phase-gratings given the non-ideal shapes. Our method measures the diffracted neutron transverse momentum distributions caused by passing through the phase-gratings and determines the optimal orientation for the phase-gratings. The analogous method for optimizing the orientation of phase-gratings for X-rays is deemed crucial for obtaining high interferometer contrast \cite{miao2016universal}. We therefore expect that our technique will find fruitful applications in optimizing neutron grating interferometers.  

\section{Phase-Grating Diffraction}


\begin{figure}
\includegraphics[width=\linewidth]{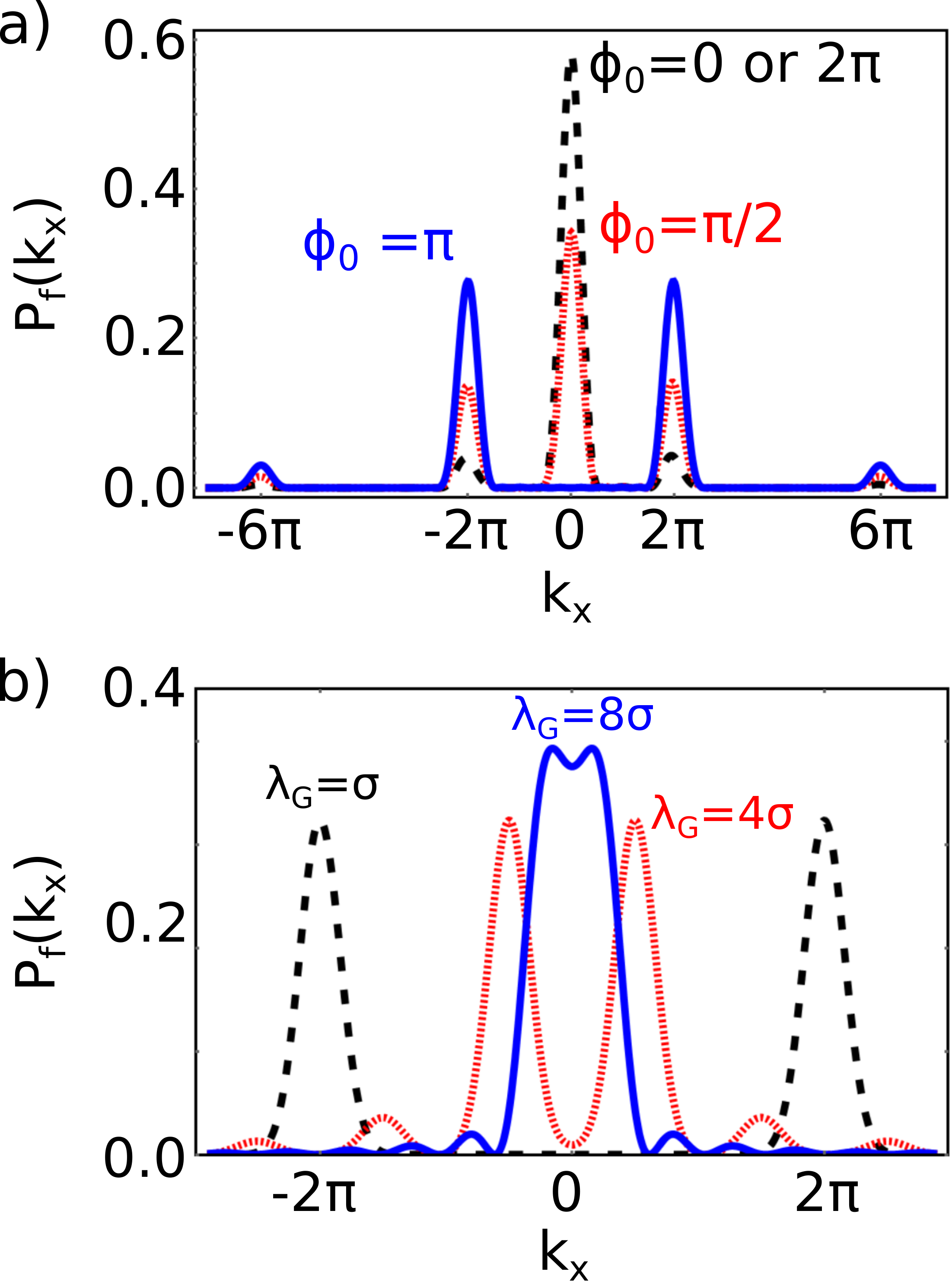}
\caption{Transverse momentum distribution of a Gaussian wavepacket with a spatial coherence length of $\sigma=1$ after passing through a $50$~\% comb-fraction phase-grating with period $\lambda_G$ and phase amplitude $\phi_0$. (a) $\lambda_G=1$ and $\phi_0$ is varied, (b) $\phi_0=\pi$ and $\lambda_G$ is varied.}
 \label{fig:sims}
\end{figure}

Neutron optics rely on refraction and diffraction phenomena for beam control and manipulation \cite{klein1983neutron}. Various shapes of materials with unique neutron indices of refraction can be constructed to induce phase shifts, diffraction, beam displacements, and spin-dependent interactions. The action of such components may intuitively be understood from the fact that inducing a spatially-dependent phase over the neutron wavefunction modifies its momentum distribution.
\par
Consider an incoming neutron wavepacket which is propagating along the $z$-direction and transversing through a material which has a thickness variation along the $x$-direction. The material induces onto the neutron wavefunction a spatially-dependent phase of $\phi(x)\sim-\lambda N b_c D(x)$, where $\lambda$ is the mean neutron wavelength, $Nb_c$ is the scattering length density, and $D(x)$ is the material thickness. 



The momentum distribution after the material is given by:

\begin{align}
P_f(k_x)=|\mathcal{F}\{e^{-i \phi(x)}\Psi_{i}(x)\}|^2,
\label{Eqn:psik}
\end{align}

\noindent where $\mathcal{F}\{...\}$ is the Fourier transform, and $\Psi_{i}(x)$ is the neutron's spatial wavefunction before the material.

If a neutron wavepacket passes through a phase-grating of period $\lambda_G$, then $P_f(k_x)$ will exhibit diffraction peaks with angular separations of

\begin{align}
\theta=\sin^{-1} \left(\frac{n\lambda}{\lambda_G}\right),
\label{Eqn:thetaquantum}
\end{align}

\noindent
where $n$ is an integer that represents the diffraction order.  The quality of the diffraction peaks depends on the spatial phase profile induced by the phase-grating.  For example, the phase profile of a $50$~$\%$ comb-fraction square wave phase-grating is given by

\begin{align}
\phi(x) = \frac{\phi_0}{2} \text{sgn}\left[\cos\left(\frac{2\pi x}{\lambda_G}\right)\right].
\label{Eqn:hf}
\end{align}

\noindent
The corresponding diffracted momentum distribution is shown in Fig.~\ref{fig:sims}a for various amplitudes $\phi_0$. Note that for a $\phi_0=\pi$ phase-grating the zeroth diffraction order is suppressed.

To resolve the individual diffraction peaks, the wavepacket's transverse coherence length $\sigma$ must be suitably long compared to the grating period.  Fig.~\ref{fig:sims}b shows $P_f(k_x)$ for a $\phi_0=\pi$ square wave phase-gratings with various periods. It can be seen that as the coherence length of the neutron wavepacket becomes shorter than the period of the grating the diffraction peaks will no longer be well-defined.

\section{Experimental  Methods}

\begin{figure}
\centering
\includegraphics[width=.9\linewidth]{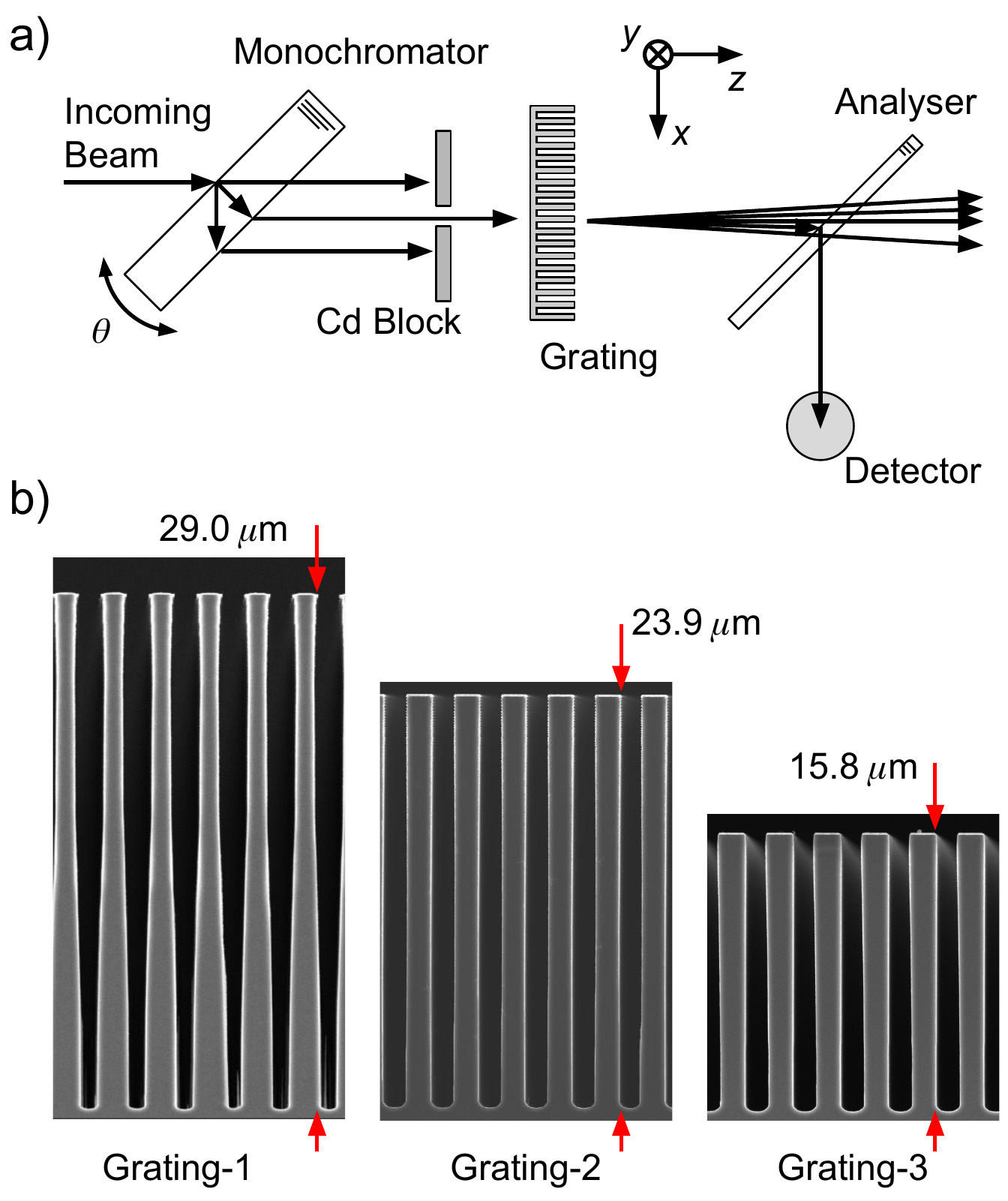}
\caption{(a) Experimental setup. A $\lambda=4.4$~\AA \; neutron beam is prepared by passing through a monochromator crystal and cadmium block.  The beam then passes through a phase-grating which induces a spatially dependent phase across the neutron wavefunction. The resulting change in the momentum distribution is measured by an analyzer crystal and a proportional counter.  The grating was rotated around all three coordinate directions. (b) SEM micrographs of the three gratings that were used in the experiment. Part (b) reproduced from Ref \cite{heacock2018sub}.}
 \label{fig:setup}
\end{figure}

The experiment was performed at the NIOFa beamline at the National Institute of Standards and Technology (NIST) Center for Neutron Research (NCNR) in Gaithersburg, MD \cite{shahi2016new, pushin2015neutron}. Fig.~\ref{fig:setup}a shows a schematic of the experiment where the neutron momentum distribution is mapped by changing the rotational alignment of the monochromator and analyzer crystal in a double crystal diffractometer (DCD). The apparatus was first introduced in \cite{heacock2018sub}, where we demonstrated a phase-recovered tomography algorithm.  An alternative method of characterizing neutron diffraction from phase-gratings consists of measuring diffraction peaks in the far field using an array of slits, as was done for cold neutrons \cite{eder1991diffraction}.  In our setup, the smaller slit widths and larger distances between slits that would be required for thermal neutrons are avoided by using a DCD.

The transverse coherence length prepared by the Bragg-diffracted beam in this setup is around $35 \; \mu \mathrm{m}$ \cite{heacock2018sub}. This is much larger than the grating period of $2.4 \; \mu \mathrm{m}$, which is a requirement to be able to resolve the grating-diffracted peaks.


Three Si gratings, all with period $\lambda_G=2.4$~$\mu$m, were analyzed. The depth of Grating-1 was $h = 29.0~\mu$m, Grating-2: $h = 23.9~\mu$m, and Grating-3: $h = 15.8~\mu$m. These thicknesses of Si correspond to a phase shift of 2.6~rad, 2.2~rad, and 1.4~rad, respectively, for $\lambda = 4.4 \; \mathrm{\AA}$ neutrons. In Fig.~\ref{fig:setup}b are the profiles of the phase-gratings as shown by the Scanning Electron Microscope (SEM) micrographs.

The gratings were first all analyzed separately.  Momentum distributions were measured as a function of rotation about three axes (defined in Fig.~\ref{fig:setup}a): (1) the axis perpendicular to the neutron propagation and grating diffraction directions ($y-$axis), (2) the axis defined by the grating diffraction direction ($x-$axis), and (3) the propagation beam axis ($z-$axis).  In addition, momentum distributions were measured with two gratings separated by $20$~cm as a function of rotation about the $z$-axis. 




\section{Rotational Effects on the Resulting Momentum Distribution}

\subsubsection*{$y$-axis rotations}

Rotations about the $y$-axis cause the effective grating period to decrease.  However, the high aspect ratio of these gratings causes a change in the phase path integral 
to be the dominant effect.  Integrating over the neutron's trajectory through a phase-grating as a function of rotation about the $y$-axis modifies the effective phase profile from a nominal square wave.  We were able to show that this data can be used to reconstruct real-space images of the grating using a phase-recovered tomography algorithm \cite{heacock2018sub}.  In this work we focus on predicting the momentum distribution from the SEM micrographs and analyzing the impact of the grating shape on interferometer performance.  Note that this effect is not as pronounced for very cold neutrons \cite{eder1991diffraction}, because the phase-gratings used for cold neutrons have aspect ratios close to unity.



\begin{figure}[t!]
\centering
\includegraphics[width=\linewidth]{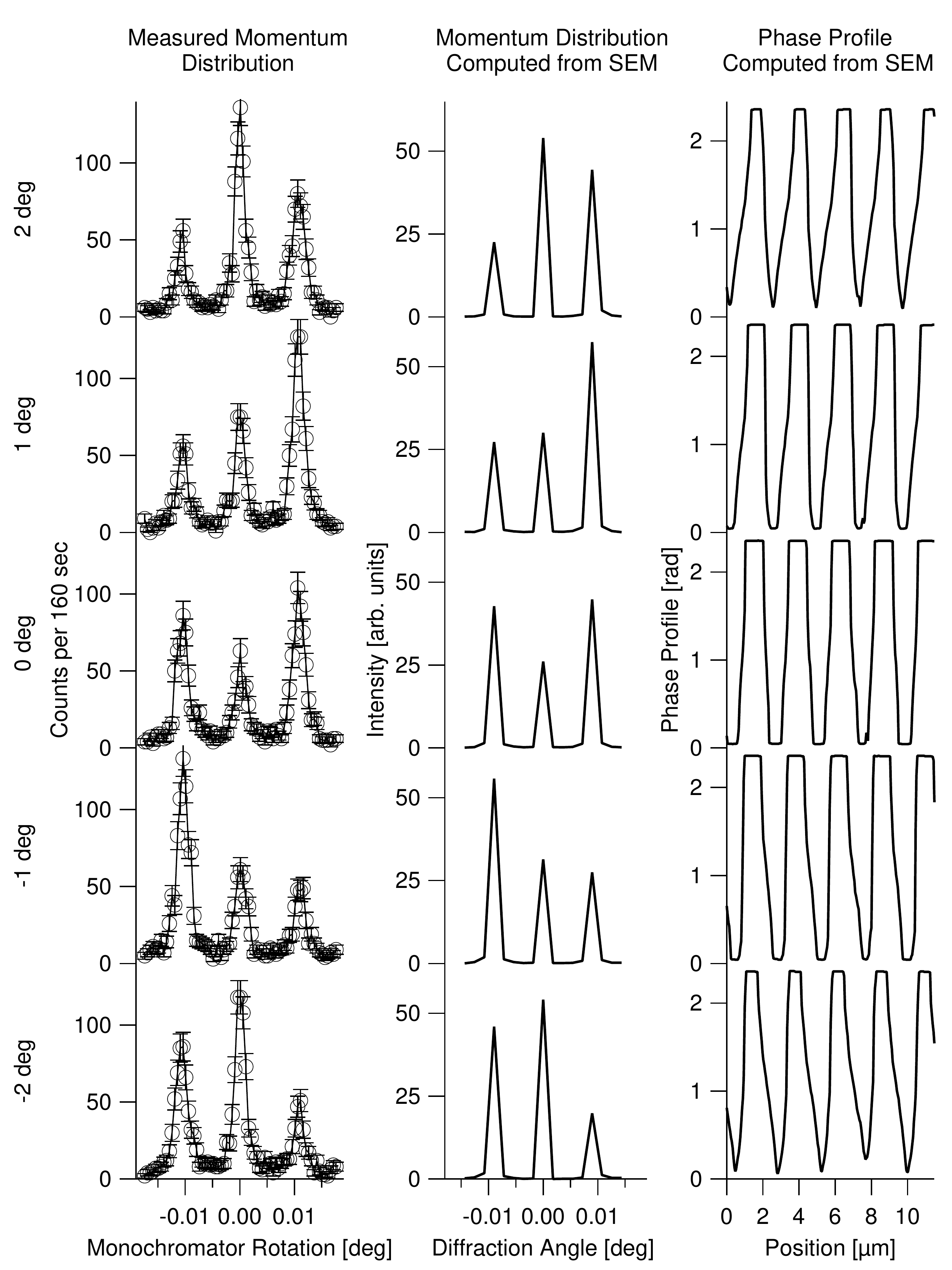}
\caption{The left column of plots is the measured momentum distributions as a function of grating rotation about the $y$-axis for Grating-1.  Grating rotation ranges from -2~degrees to 2~degrees in one degree steps from bottom to top. Uncertainties are purely statistical, and lines connecting data points are shown for clarity. The middle (right) column shows the diffracted momentum distributions (phase profiles) computed from the SEM micrographs.}
\label{fig:yRotSpec}
\end{figure}

To quantify this effect, the effective phase profiles were computed from the SEM micrographs as a function of rotation about the $y$-axis 
using silicon's neutron scattering length density of $2.1 \times 10^{-6}$ \AA$^{-2}$ \cite{NCNRweb}.  The  momentum distributions were computed by transforming the phase profiles taken from the SEM micrographs according to Eq.~\ref{Eqn:psik} using a fast Fourier transform (FFT).  The power spectra of the FFT compared to experimental momentum space distributions for Grating-1, along with the effective phase profiles, are shown in Fig.~\ref{fig:yRotSpec}.

\begin{figure}[t!]
\centering
\includegraphics[width=0.8\columnwidth]{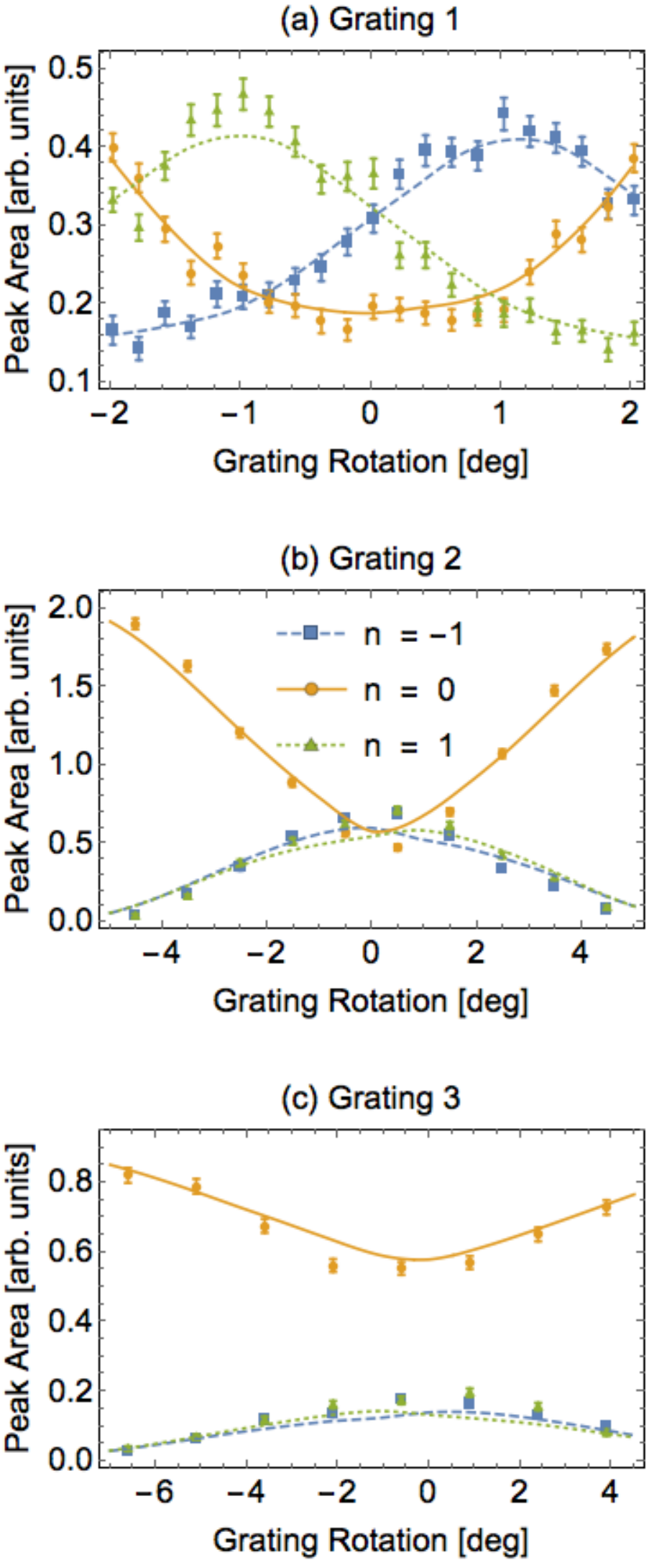}
\caption{Zeroth and first order diffraction peak areas from the measured momentum distributions (data points) and the diffracted momentum distributions computed from the SEM micrograhs (solid lines) as a function of grating rotation about the $y$-axis for (a) Grating-1, (b) Grating-2, and (c) Grating-3. Uncertainties shown are purely statistical.}
\label{fig:roty}
\end{figure}

The diffraction peaks were characterized by fitting the measured intensity versus monochromator rotation to multiple Lorentzians.  Fig.~\ref{fig:roty} compares the areas of the fitted Lorentzians to the areas of the peaks in the FFT power spectra computed from the SEM micrographs.  The computed diffraction peak areas from the SEM micrographs were scaled to match the absolute peak areas of the measured distributions, with the relative areas between diffraction orders and grating rotations preserved. The asymmetry in the area under the two first-order diffraction peak areas was unique to Grating-1.  This effect can be understood by inspecting the spatial phase profiles computed from the SEM micrographs (see Fig~\ref{fig:yRotSpec}).  As the grating is rotated, the phase profile has an asymmetrical slope.  The sign of the slope changes as the grating rotation crosses zero. This unique phase profile is only seen in Grating-1 because of the trapezoidal grating walls which are visible in the SEM micrograph of Fig.~\ref{fig:setup}b.

Due to the high aspect ratio of the phase-gratings, the spatially varying phase profile vanishes as the rotation about the $y$-axis approaches $ \lambda_G/h \sim 5$~degrees.  The quality of the grating further diminishes the angular range over which the grating is effective.  This can be seen by inspecting the relative peak areas in Fig.~\ref{fig:roty}.  Because of the distorted walls of Grating-1, the first order peak areas have a large slope as a function of $y$-axis rotation when the grating is aligned to the beam.  This adds a transverse momentum dependence to the diffracted wave amplitudes if the grating is used in a phase-grating NI, which diminishes interferometer contrast.  Even for the typical beam divergence of $\sim 0.5$~degrees, the angular edges of the beam probe drastically different grating profiles.  This effect is absent in Grating-2 and Grating-3, where the diffracted peak areas form a maximum when the grating is aligned to the beam.

\begin{figure}[t!]
\centering
\includegraphics[width=0.85\columnwidth]{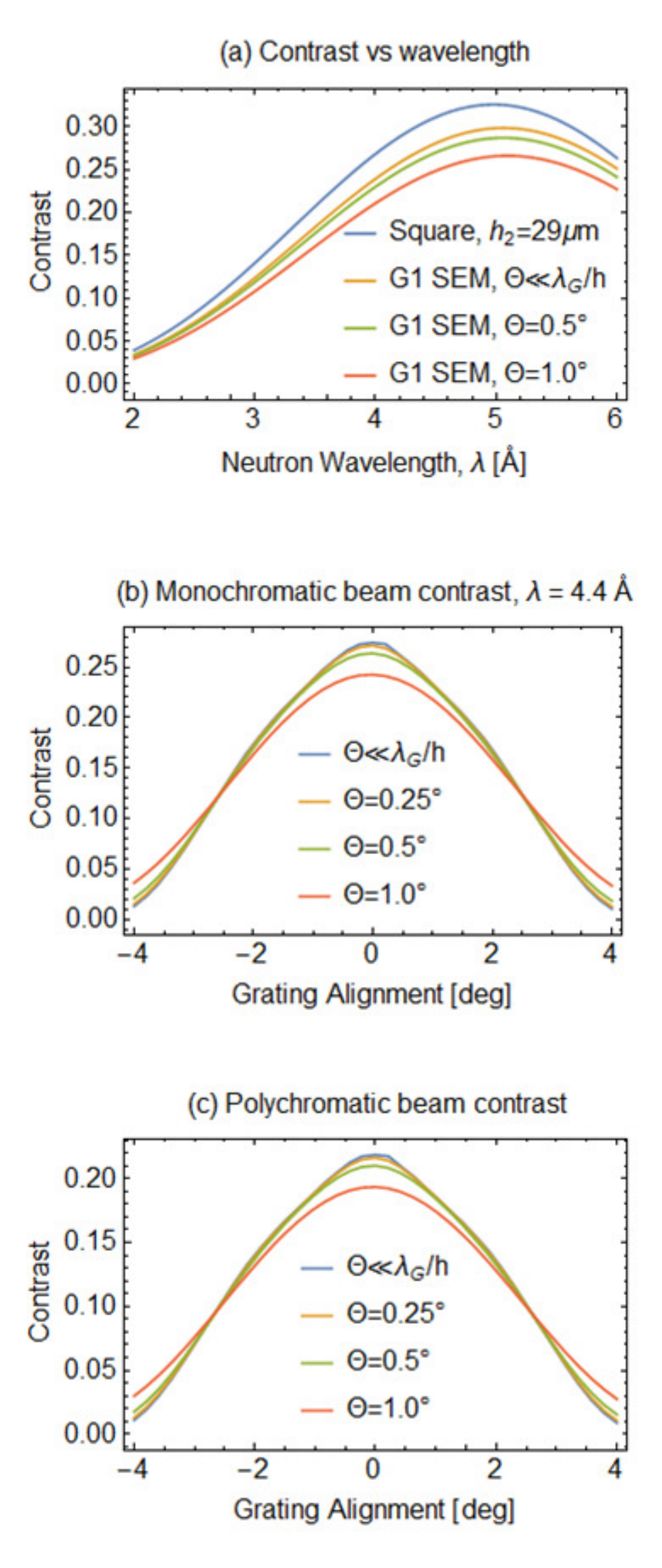}
\caption{Predicted contrast of a three phase-grating \moire interferometer, with the first and third gratings ideal square gratings with heights 15~$\mu$m and 16~$\mu$m respectively, and the central grating given by the SEM of Grating-1.  (a) Contrast versus wavelength with all gratings aligned to the beam and beam divergence of $\Theta$. (b-c) Contrast versus central grating angular alignment about the $y$-axis. For (b) $\lambda = 4.4$~\AA \; and for (c) $\lambda$ is given by a Maxwell-Boltzmann distribution with $\lambda_c=5$~\AA\;.  Also shown in (a) is the ideal contrast for a square grating of height $29 \; \mu \mathrm{m}$.  Note the loss of contrast for all wavelengths given the non-square shape of Grating-1.}
\label{fig:InfCon}
\end{figure}

The impact of a non-ideal grating profile on the contrast of a three phase-grating \moire interferometer is is illustrated in Fig.~\ref{fig:InfCon}.  Maximum achievable contrast was calculated by computing the diffracted wave amplitudes from the Grating-1 SEM as previously described.  This was combined with the closed-form solution for the diffracted wave amplitudes for square gratings with heights of $15~\mu \mathrm{m}$ and 16~$\mu$m.  The three phase-grating \moire pattern maximum contrast was then computed according to the equations set out in \cite{miao2016universal}, with Grating-1 used as the central grating.  Contrast is computed as a function of wavelength and the angular alignment of the central grating about the $y$-axis. The resulting function can then be integrated over a wavelength and/or angular distribution. In our case, we take the angular distribution to be Gaussian with width $\Theta$ and assume either a monochromatic or Maxwell-Boltzmann distributed source.  A reduction in contrast from the non-square profile of Grating-1 is clearly visible for all wavelengths (Fig.~\ref{fig:InfCon}a), and beam divergence at the level of most neutron experiments ($\sim 0.5 ^\circ$) also reduces contrast.  Note that the effect of beam divergence is only considered for the central grating, which would dominate the effect, as it is twice the height of the other two gratings.  Finally, computing contrast versus the angular alignment of the central grating for a neutron beam with a wavelength distribution given by a Maxwell-Boltzmann distribution with $\lambda_c = 5$~\AA, as was the case in the first demonstration of a neutron three phase-grating \moire interferometer \cite{sarenac2018three}, we show that the lower-than-expected contrast could have occurred from a 3.5~degree misalignment of the central grating about the $y$-axis.

\subsubsection*{$x$-axis Rotations}

Rotations about the $x$-axis only change the effective thickness of the grating, and therefore the phase profile of the grating changes as 

\begin{equation}
\phi(x) \rightarrow {\phi(x) \over \cos \alpha} .
\label{eqn:yrot}
\end{equation}

\begin{figure}
\includegraphics[width=0.9\linewidth]{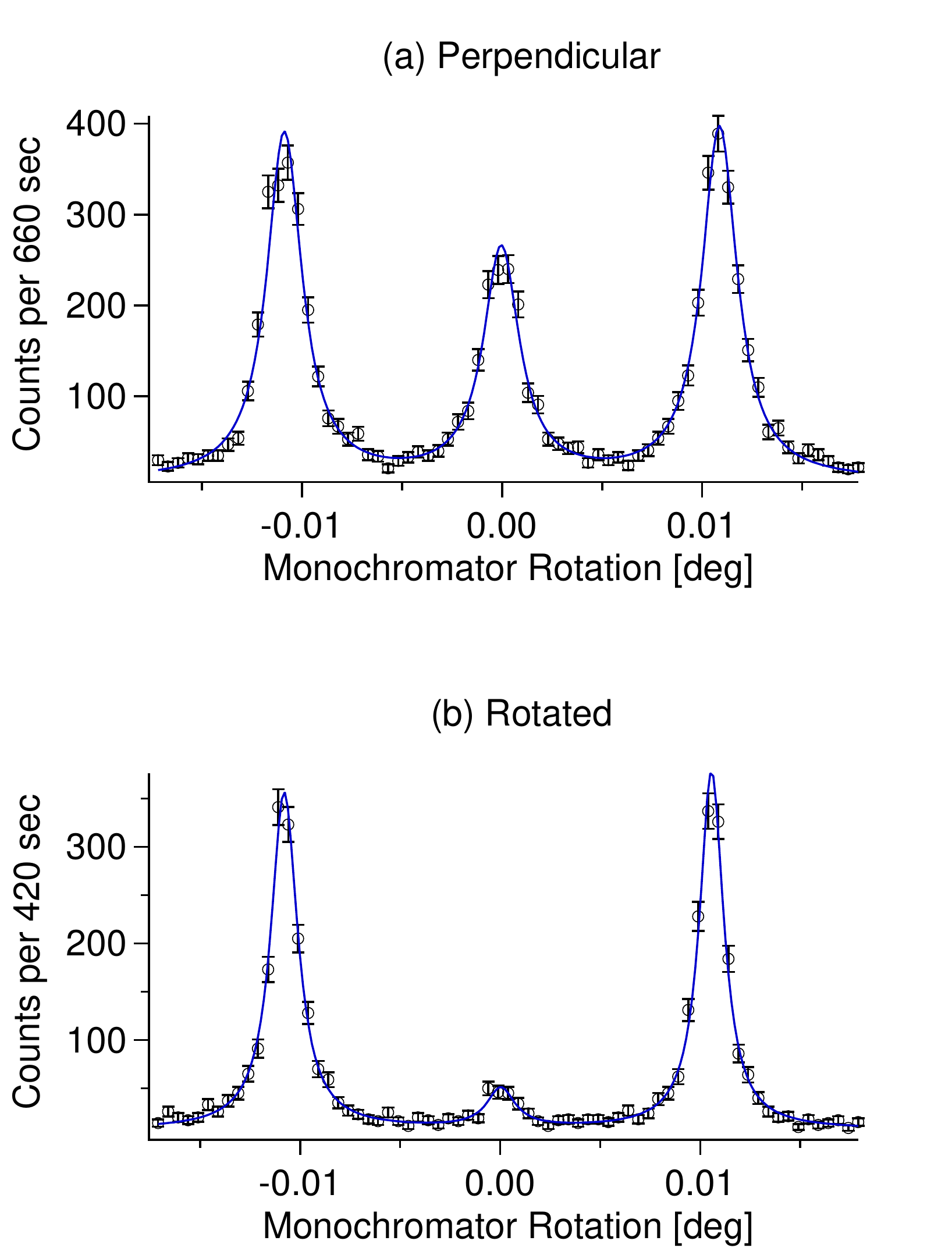}
\caption{Grating-2 diffraction peaks when the grating is (a) perpendicular to the incoming beam and (b) rotated by $\sim 40$~degrees about the $x$-axis so that it is approximately acts as a $\pi$ phase-grating. Uncertainties shown are purely statistical.  Lines are fitted Lorentizians.}
\label{fig:rotx}
\end{figure}

\noindent
A $50$~$\%$ comb-fraction grating with a $\pi$ phase amplitude theoretically has the zeroth-order peak suppressed as shown in Fig.~\ref{fig:sims}a.  As the amplitude of the zeroth order peak is proportional to $\cos^2 (\phi_0 / 2)$, the peak reappears slowly as a function of grating thickness.  Demonstration of the zeroth-order peak suppression is shown in Fig.~\ref{fig:rotx} by rotating Grating-2 about the $x$-axis.  

However, the disappearance of the zeroth-order peak does not hold true for a $\pi$ phase-grating with sloped walls.  Thus, if a $\pi$ phase-grating is known to be aligned to the $y$-axis, as in the previous section, the lack of a zeroth order peak is a measure of the squareness of the grating.  The almost complete suppression of the zeroth order peak seen in Fig.~\ref{fig:rotx}b matches the SEM micrograph in Fig.~\ref{fig:setup} in that Grating-2 is of very high uniformity and its shape closely estimates a square wave.

\subsubsection*{$z$-axis rotations}

\begin{figure}
\includegraphics[width=0.95\linewidth]{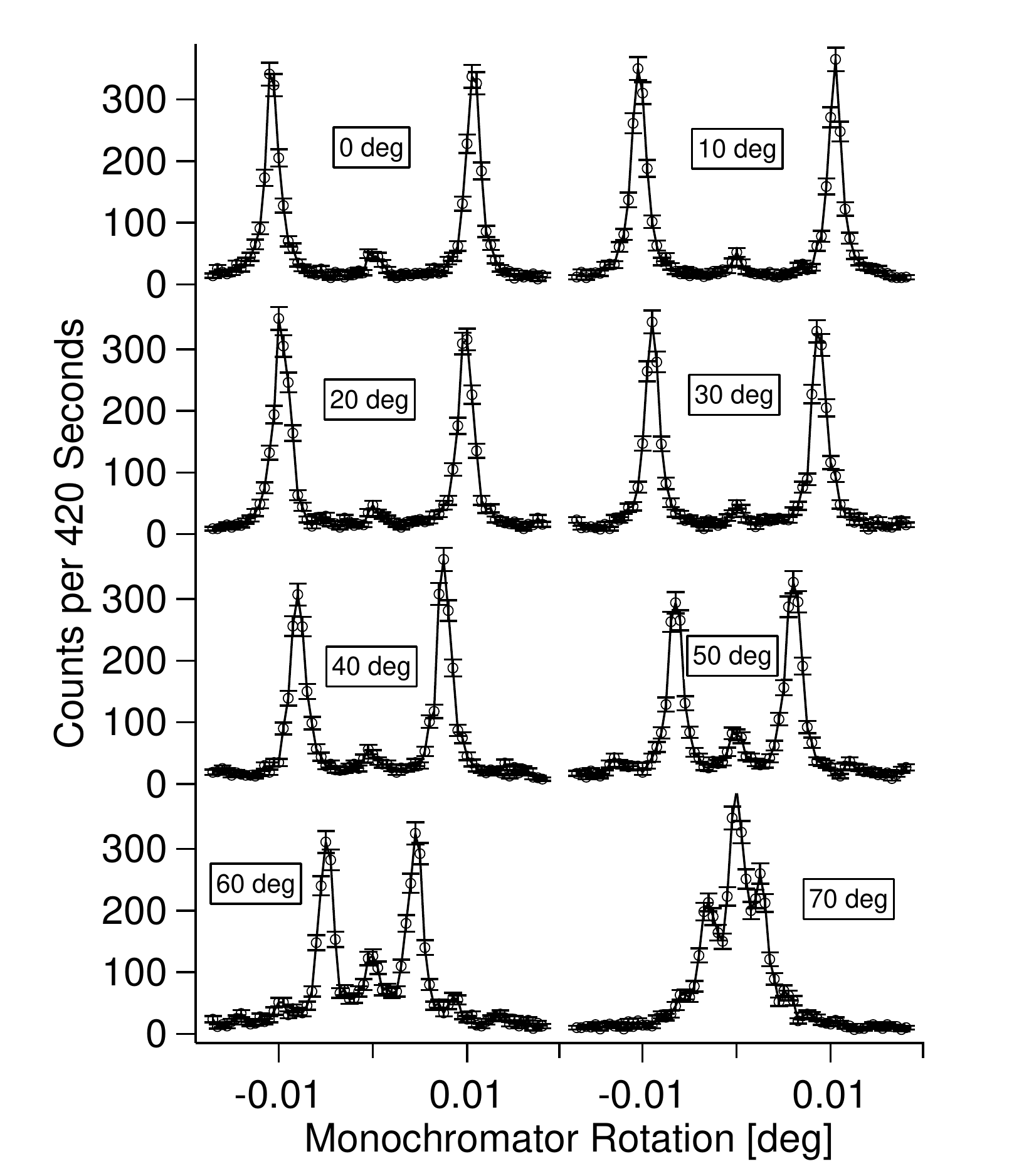}
\caption{Grating-2 diffraction peaks in the $\sim \pi$ orientation as a function of rotation about the $z$-axis.  Uncertainties shown are purely statistical.  Lines connecting data points are shown for clarity.}
\label{fig:rotz}
\end{figure}

Grating rotations about the $z$-axis rotate the diffraction plane of the grating out of the diffraction plane of the monochromator-analyzer pair (the $x$-$z$ plane in Fig.~\ref{fig:setup}a). This can be interpreted as effectively lengthening the grating's period by 

\begin{equation}
\lambda_G \rightarrow {\lambda_G \over \cos \gamma} ,
\label{eqn:pcos}
\end{equation}

\noindent
where $\gamma$ is the angle of elevation between the crystal and grating diffraction planes about the $z$-axis. Because the momentum distribution is only narrow in the crystal diffraction direction, rotating the grating about the $z$-axis reduces the $\sigma/\lambda_G$ ratio  (see Fig.~\ref{fig:sims}b).  This causes the measured diffraction peaks to overlap as the grating is rotated as shown in Fig.~\ref{fig:rotz}.

\subsubsection*{z-axis rotations with two gratings}



\begin{figure}
\includegraphics[width=0.9\linewidth]{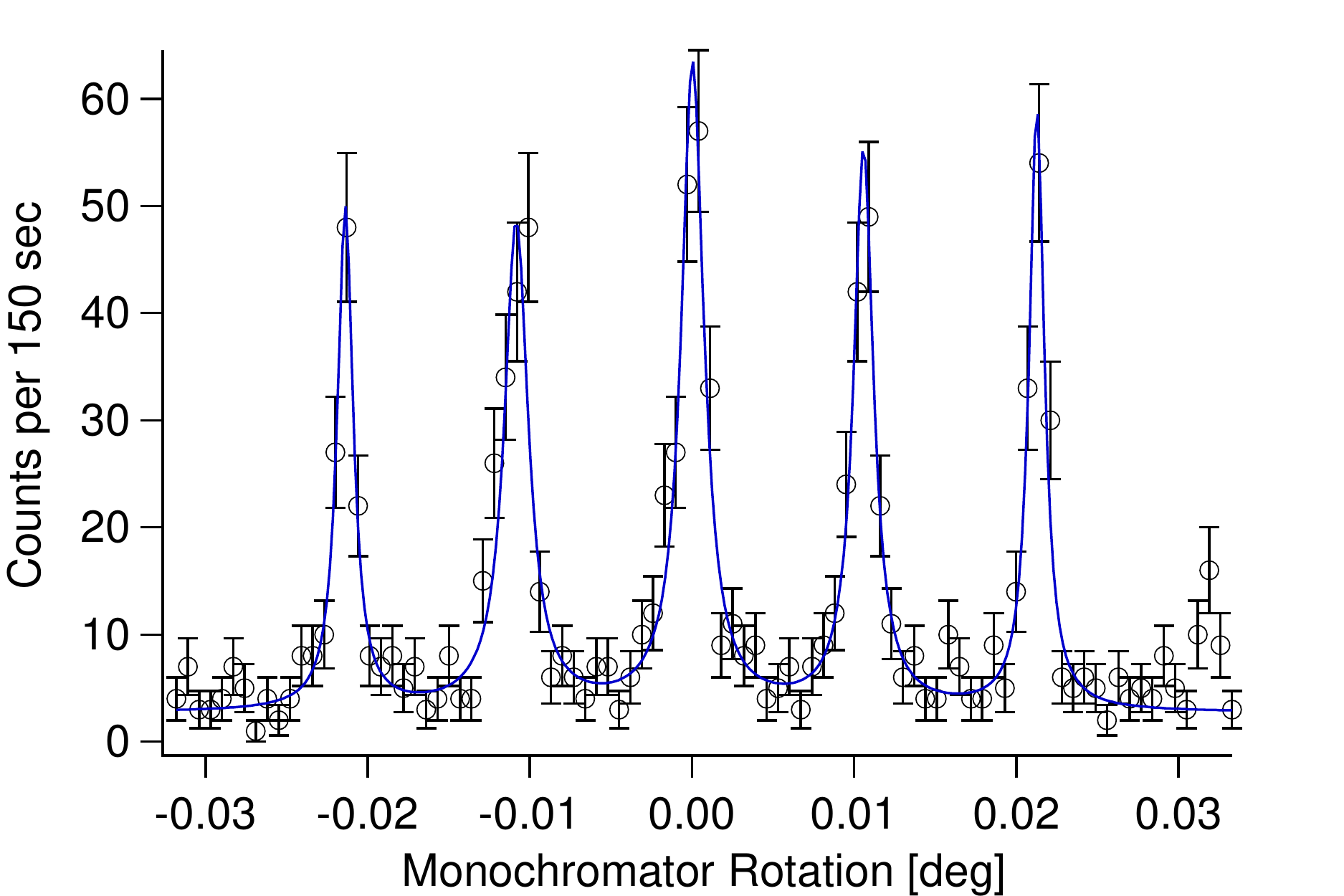}
\caption{Double grating momentum distribution with Grating-1 and Grating-2 and fitted Lorentzians.  Each grating was aligned to the beam about the $y$-axis separately.  Uncertainties shown are purely statistical.}
\label{fig:twoG}
\end{figure}

When two gratings with the same period are used in series, the $n^{th}$-order diffraction peaks from the first grating are diffracted into $(n+m)^{th}$ order peaks, where $m$ is the diffraction order of the second grating.  
The measured momentum distribution of Grating-1 and Grating-2 operated in series is shown in Fig.~\ref{fig:twoG}.  Second order diffraction peaks now have amplitudes similar to the first and zeroth order peaks.

\begin{figure}
\includegraphics[width=0.9\linewidth]{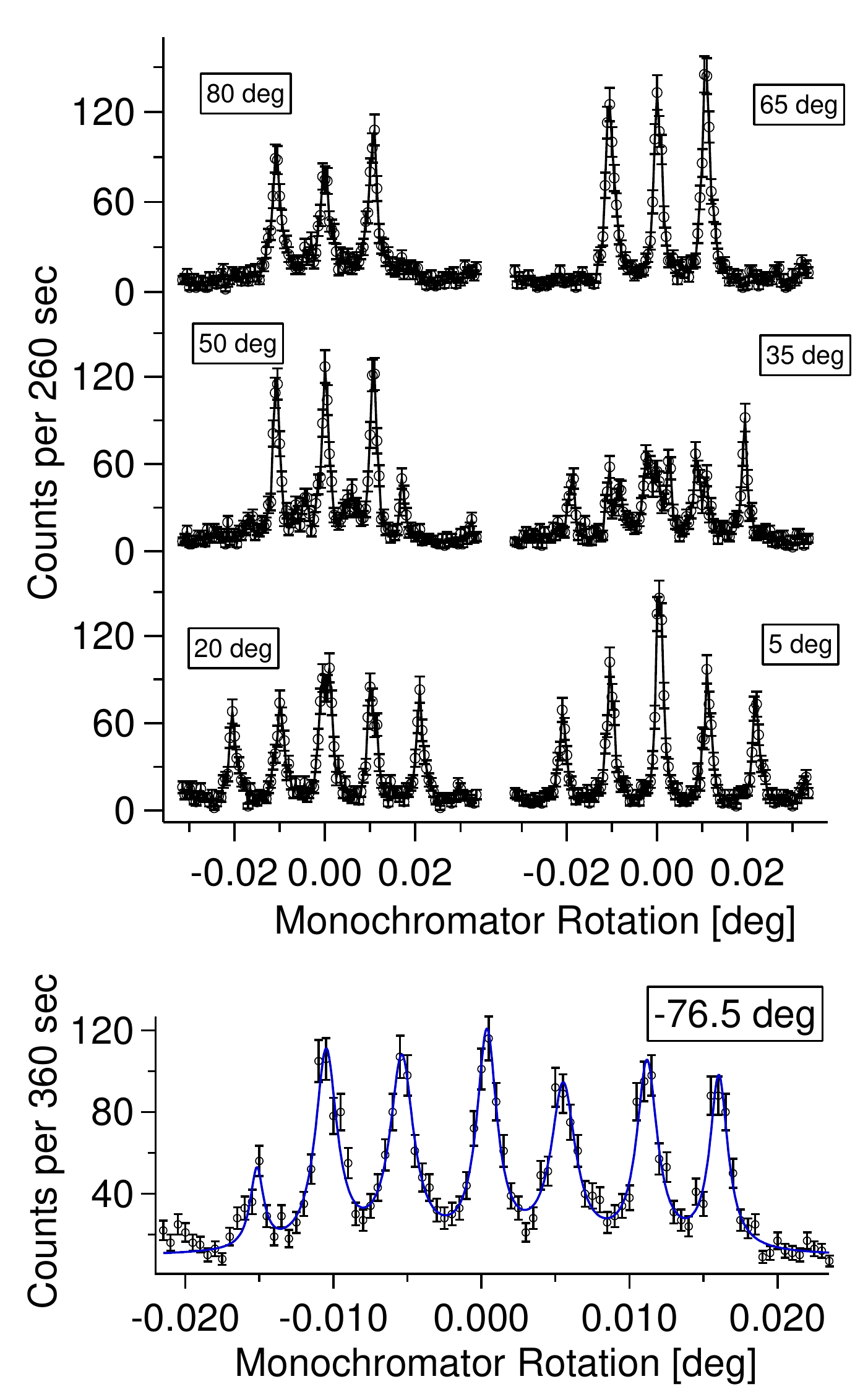}
\caption{Two grating measured momentum distributions with Gratings-1 and -2.  Grating-1 was rotated about the $z$-axis while Grating-2 was held fixed.  Each grating was aligned to the beam about the $y$-axis separately.  Uncertainties shown are purely statistical.  In the upper array of plots, lines between data points are shown for clarity.  Fitted Lorentzians are shown in the bottom plot.}
\label{fig:twoGy}
\end{figure}

If one or both of the gratings are rotated about the $z$-axis, Eqn.~\ref{Eqn:thetaquantum} is modified and the crystal monochromator / analyzer pair will measure peak locations of

\begin{equation}
\theta = \sin^{-1}\left [ {\lambda \over \lambda_G} (n \cos \gamma_1 + m \cos \gamma_2 ) \right ]
\label{eqn:peakpos}
\end{equation}

\noindent
for two gratings of the same period $\lambda_G$, with $\gamma_1$ and $\gamma_2$ the $z$-axis rotation of Grating-1 and Grating-2, respectively.  The effect of rotating one grating about the $z$-axis while leaving the other aligned is shown in Fig.~\ref{fig:twoGy}.  The number of observable peaks increases because $n + m \cos\gamma $ need not add to an integer between $-2$ and $2$ for $n$ and $m$ restricted to $\{ 0, \pm 1 \}$.  As shown in Fig.~\ref{fig:twoGy}b seven diffraction peaks were resolvable; though up to nine peaks are possible.

\section{Conclusion}

We have demonstrated a method to determine the grating orientation which produces the optimal grating interferometer contrast. By characterizing the diffraction peaks of a neutron phase-grating with a crystal monochromator-analyzer pair, we show how the grating's alignment with the beam impacts the resulting momentum distribution.  Rotations about each of the three axes for the coordinate system defined by the diffraction plane of the grating resulted in different effects.  

Rotations about the $y-$axis change the effective phase profile of the grating.  This alters the relative diffracted wave amplitudes in a way that can be predicted from the SEM micrographs of the grating.  We find that a grating with distorted walls exhibits a gradient in the measured first order diffraction peak areas as a function of rotation about the $y$-axis, which degrades contrast in a phase-grating NI. The overall phase of the grating is set by its depth and the neutron wavelength; rotating the grating about the $x$-axis changes the effective depth of the grating.  We used this degree of freedom to tune a grating to have a phase of $\phi_0 = \pi$ and almost completely suppress the zeroth order diffraction peak. Rotations about the $z$-axis tilts the diffraction plane of the grating relative to the monochromator-analyzer pair.  This allowed us to measure the outgoing momentum distribution as the grating period approached and became larger than the coherence length of the neutron wavepacket.

Measuring the momentum distribution of a neutron beam diffracted by a phase-grating provides a useful grating characterization tool.  A crystal monochromator-analyzer pair could be used in-situ for the pre-alignment of grating NIs, where each grating could be aligned to the incoming beam separately.  For high contrast in a three phase-grating \moire interferometer, the middle grating needs to act as a good $\pi$ phase-grating \cite{miao2016universal}. Here we can conclude that the demonstration of the neutron three phase-grating \moire interferometer achieved a low contrast of 3\% most likely because it used the trapezoidal-shaped Grating-1, without any rotational optimization, as the middle grating. 

\section{Acknowledgments}

This work was supported by the U.S. Department of Commerce, the NIST Radiation and Physics Division, the Director's office of NIST, the NIST Center for Neutron Research, the National Institute of Standards and Technology (NIST) Quantum Information Program, the US Department of Energy under Grant No. DE-FG02-97ER41042, and National Science Foundation Grant No. PHY-1307426. This work was also supported by the Canadian Excellence Research Chairs (CERC) program, the Canada  First  Research  Excellence  Fund  (CFREF), the Natural Sciences and Engineering Research Council of Canada (NSERC) Discovery program, and the Collaborative Research and Training Experience (CREATE) program.

\bibliography{lib}

\end{document}